\documentclass[a4paper,11pt]{article}

\usepackage{pos}
\renewcommand{\logo}{\relax}

\DeclareUnicodeCharacter{223C}{\textasciitilde}
\DeclareUnicodeCharacter{22EF}{\ldots}
\DeclareUnicodeCharacter{0229}{\c{e}}
\usepackage[utf8]{inputenc}

\usepackage{graphicx}
\usepackage{dcolumn}
\usepackage{bm}
\usepackage{hyperref}

\usepackage{xurl}
\usepackage{xcolor}

\newcommand{\pbar}{\bar{p}}
\newcommand{\nbar}{\bar{n}}

\let\OLDthebibliography\thebibliography
\renewcommand\thebibliography[1]{
  \OLDthebibliography{#1}
  \setlength{\parskip}{3pt}
  \setlength{\itemsep}{3pt}
}

\title{Precision Spectroscopy of Antiprotonic Atoms for Investigation of Low-energy Antinucleon--nucleus Interactions}
\ShortTitle{Precision Spectroscopy of ...  Low-energy Antinucleon--nucleus Interaction}

\author*[a]{Takashi Higuchi}
\author[b]{Hiroyuki Fujioka}

\affiliation[a]{Institute for Integrated Radiation and Nuclear Science, Kyoto University\\
2-1010 Asashiro-nishi, Kumatori, Osaka, Japan
}

\affiliation[b]{ Department of Physics, Institute of Science Tokyo\\
2-12-1 Ookayama, Meguro, Tokyo, Japan}

\emailAdd{higuchi.takashi.8k@kyoto-u.ac.jp}
\emailAdd{fujioka@phys.sci.isct.ac.jp}

\abstract{
\textbf{Abstract:}
We propose a high-precision x-ray spectroscopy experiment of antiprotonic atoms to advance the understanding of low-energy antinucleon--nucleus interactions. The current leading model of antiproton--nucleus interactions is based  on an optical potential with parameters derived from a global fit to antiprotonic atom x-ray data across the periodic table. However, the isovector parameter of this potential remains poorly constrained due to uncertainties in   nucleon distributions of the nuclei. To address this, we propose to use calcium isotopes with well-studied nucleon distributions to minimize these uncertainties. 
A superconducting microcalorimeter detector will provide a resolution of 50--70~eV in the energy range of interest, allowing high precision determination of the isotope-dependent strong-interaction shifts and widths.
The outcomes of the proposed experiment  can be used to  refine the model of antinucleon--nucleus interactions and provide critical data for  future experiments searching for neutron--antineutron oscillations.
}

\FullConference{%
  International Conference on Exotic Atoms and Related Topics and Conference on Low Energy Antiprotons
(EXA-LEAP2024)\\
  26--30 August, 2024\\
 Austrian Academy of Sciences, Vienna
}

\begin{document}
\maketitle

\section{Background --- studies of hadronic interactions with antiprotonic atoms}
Spectroscopy of exotic atoms plays an important role in studying the properties of their constituents and their fundamental interactions. Among these, hadronic atoms, in which a negatively charged hadronic particle such as $\pi^-$, $K^-$, or $\bar{p}$ replaces an electron of the atom and is bound to the nucleus, provide information about the strong interaction that complements other experimental techniques. The major experimental observables in  hadronic atom  spectroscopy are the energies and broadening of atomic  transitions of the orbiting  particle. These measurements can be compared with calculations to extract  the strong-interaction shifts and widths   induced by the hadron--nucleus interaction. The formalism of the optical potential that represents such low-energy hadron-nucleus interaction has been established  through previous systematic studies of hadronic atoms~\cite{Batty1997}. In case of $\pbar$ atoms, the  $s$-wave potential $V_{\rm opt}$ in the following form is commonly employed,
\begin{equation}\label{eqn:Vopt}
    2\mu V_{\mathrm{opt}}(r)=-4\pi\left(1+\frac{\mu}{M}\frac{A-1}{A}\right)[(b_0(\rho_n (r) +\rho_p(r) )+b_1(\rho_n(r) -\rho_p(r) )].
\end{equation}
Here, $\mu$ is the $\pbar$--nucleus reduced mass,  and $M$ the nucleon mass. $\rho_n(r)$ and  $\rho_p(r)$ are the neutron and the proton density distributions normalized to the number of neutrons $N$ and the number of protons $Z$, respectively. $A=N+Z$ is the mass number of the nucleus. Complex parameters $b_0$ and $b_1$ are to be determined by an analysis,  corresponding  respectively to the isoscalar and isovector $\pbar$--nucleon scattering lengths in the impulse approximation~\cite{Friedman2007}.

The latest theoretical model is based on an analysis by Friedman \textit{et al.}~\cite{Friedman2005} of systematic data obtained by the PS209 experiment at LEAR, consisting of  90 data points of strong-interaction shifts and widths in  $\pbar$ atoms and 17 data points of radiochemical measurements~\cite{Trzcinska2001}. The optical potential of Eq. (\ref{eqn:Vopt}) was fitted to the experimental data across the periodic table to determine  $b_0$, $b_1$ together   with nucleon  density distribution parameters and a finite-range parameter.  The  obtained results are consistent with $b_1=0\,{\rm fm}$ and $b_0= 1.0\,(1)+ 1.3\,(1) {\rm i} \,{\rm fm} $ with a Gaussian-folding parameter of  $0.85\,{\rm fm}$ to  account for a finite interaction range~\cite{Friedman2005}. 
This optical potential reproduced   data of $\pbar$--nucleus elastic scattering experiments with the parameters consistent with the $\pbar$ atom global fit,  validating this  potential  across the energy threshold~\cite{Janouin1986,Friedman2014}.

\section{Motivations for further investigation of the antinucleon--nucleus interactions}
Although the optical potential  model is successful in reproducing  data of both atomic and scattering experiments, 
several factors motivate further investigation of antinucleon--nucleus interactions at low energies. 

Firstly, there is a discrepancy  in antinucleon--nucleus annihilation cross sections, as  pointed out by Friedman~\cite{Friedman2014,Friedman2015c}. 
As done for the elastic scattering analysis,  
the optical potential derived from the $\pbar$ atom global fit can be used to calculate $\pbar$-- and $\nbar$--nucleus annihilation cross sections~\cite{Friedman2014}. The obtained results showed severe discrepancies from  $\nbar$--nucleus annihilation cross-section data obtained by the OBELIX experiment~\cite{Astrua2002}. 
Notably, the theory could not reproduce an increase in the experimental annihilation cross sections at low energies, underestimating them by factors of 2 to 4 at the lowest measured momentum~\cite{Friedman2014}. 
This apparent puzzle has triggered various theoretical and experimental investigations in recent years, including comparative studies with a model based  on a fundamental $N\bar{N}$ interaction potential and alternative parametrization of nucleon density distributions~\cite{Friedman2015c,Lee2018,Protasov2020}. On the experimental side, efforts have been made to measure the $\pbar$--nucleus annihilation cross sections at momenta of 100 MeV/c and below, which can be compared with the OBELIX $\nbar$ data~\cite{Bianconi2011,Aghai-Khozani2018,Aghai-Khozani2021}. However, the issue still remains unresolved.

Secondly, there has been growing interest in low-energy antinucleon--nucleus interactions due to their relevance in experiments searching for neutron--antineutron ($n$--$\nbar$) oscillations. The $n$--$\nbar$ oscillations, which violate  baryon number by 2 while conserving  lepton number,  are unique channels to probe  physics beyond the Standard Model and test  Grand Unified  Theories~\cite{Phillips2016}. Current experimental limits on  the oscillation time are placed to be $>8.6\times10^{7}$~s (90\% C.L.)  for free neutrons and $>4.7\times10^{8}$~s (90\% C.L.) for neutrons bound in $^{16}\mathrm{O}$ nuclei~\cite{Baldo-Ceolin1994,Abe2021}.  In view of next-generation experiments proposed in future facilities, a number of  new ideas have been proposed to  significantly enhance experimental sensitivities~\cite{Nesvizhevsky2019,Gudkov2020,Protasov2020,Kerbikov2019,Shima2023}.  The essence of these proposals is to design a surface or volume with a minimal potential difference experienced by $n$ and $\nbar$, thereby suppressing the violation of degeneracy between the $n$ and $\nbar$ states.  These experiments utilize low-energy neutrons with kinetic energies below $\mathcal{O}(10)$~meV, or even as low as $\mathcal{O}(100)$~neV, where scattering phenomena are  governed by the $s$-wave scattering length. 
Due to present unavailability of a facility to use  $\nbar$ and the intrinsic difficulty to produce  $\nbar$ with such low energies,
the $\nbar$--nucleus optical potential obtained by the isospin transform of the $\pbar$--nucleus optical potential of Eq.~(\ref{eqn:Vopt}) ($n\leftrightarrow p,\,\pbar  \leftrightarrow \nbar$)  provides critical information on low-energy $\nbar$--nucleus interactions, as has already been used to derive the limit for bound neutrons~\cite{Friedman2008}. 
The precision of the $\nbar$--nucleus scattering lengths required for these proposals ranges from a few to 10\%. As these uncertainties ultimately impact the  sensitivities of  $n$--$\nbar$ oscillation searches, improvement in the precision of the antinucleon--nucleus optical potential would greatly benefit all these experimental concepts.

\section{Proposal of antiprotonic calcium atom spectroscopy with improved precision}

In this context, we propose an improved spectroscopy experiment on $\pbar$ atoms with the aim of reexamining the currently-accepted isoscalar ($b_1=0$) optical potential. 
In the previous  analysis of the global fit,  determination of $b_1$ was correlated with the nucleon  distribution parameters, which were also determined in the fit. The results could therefore vary depending on the assumptions made about the nucleon distributions~\cite{Friedman2005,Friedman2007}. Thus, the determination of $b_0$ and $b_1$ was limited by the knowledge of  nucleon distributions. Our approach here is to use nuclides   with well-studied  nucleon distributions to minimize theses uncertainties.  From several perspectives, calcium (Ca) isotopes stand out as the most suitable candidates for this study:
\begin{itemize}
    \item Calcium has  a long chain of stable isotopes, two of which are doubly magic ($^{40}\mathrm{Ca}$ and $^{48}\mathrm{Ca}$). For these nuclides, extensive studies on nucleon distributions have been conducted from both theoretical and experimental sides. The  nucleon density distributions are available from nuclear density functional theories (DFT)  as well as \textit{ab initio} methods~\cite{Naito2023,Hagen2016}, and have been measured with various experimental techniques~\cite{Zenihiro2018,Matsuzaki2021a,Adhikari2022}. 
    
    \item $^{40}$Ca has $N=Z=20$, where the contribution of $b_1$ to the optical potential is canceled to the first order. Therefore, measurements of the strong-interaction shifts and widths for other isotopes relative to $^{40}$Ca will allow for a clear extraction of the $b_1$ contribution.
    
\end{itemize}
Exploiting these characteristics of calcium isotopes, theoretical nucleon density distributions can be utilized as inputs, rather than outputs, of an analysis to extract  $b_0$ and $b_1$.  Based on this approach, Yoshimura \textit{et al.} recently reanalyzed the existing  data on $\pbar$Ca atoms from the PS209 experiment~\cite{Yoshimura2024}. In the PS209 measurement, the strong-interaction shifts of the $6h\to5g$ antiprotonic  transition were obtained for $^{40,42,43,44,48}$Ca, revealing a trend of an  increasingly repulsive shift  with a larger mass number~\cite{Hartmann2002}. The analysis, incorporating the latest theoretical nucleon density distributions as inputs, found that this trend indicates a finite value of $b_1$~\cite{Yoshimura2024}. 

\begin{figure}[tb]
    \centering
    \includegraphics[width=0.7\linewidth]{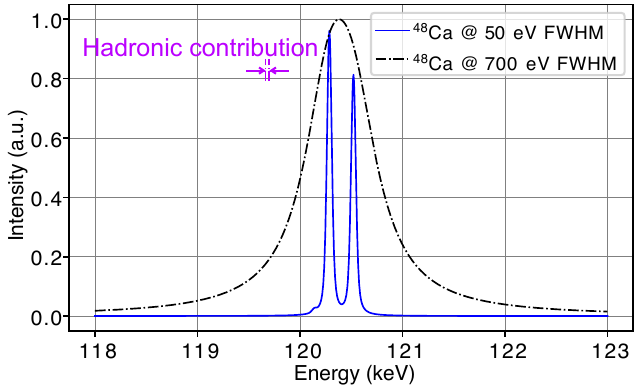}
    \caption{ Predicted  spectra  of the  $6h\to5g$ transitions of  $\pbar\,^{48}$Ca atoms. The black dashed line represents a spectrum with a  typical resolution of an HPGe detector such as used by  the  PS209 experiment~\cite{Hartmann2002}. The blue solid line represents the one expected  by a 50-eV resolution TES detector, where the  fine structure is resolved. The magnitude of the strong-interaction shift of 33~eV expected from the PS209 result  is also indicated. }
    \label{fig:pbarCa}
\end{figure}


The technological key to this proposal is the use of multi-pixel transition-edge sensor (TES) superconducting microcalorimeters, which offer high resolution ($10^{-4}$ intrinsic resolution) and high detection efficiency (0.4 quantum efficiency)~\cite{Ullom2015}. Although conventional crystal spectrometers can achieve even higher energy resolution, they come at the cost of significantly reduced detection efficiency and a limited energy range. With both high resolution and high detection efficiency, TES technology enables high-precision x-ray spectroscopy of exotic atoms, particularly in cases where luminosity is limited~\cite{Hashimoto2022,Okumura2023}. In  the AD/ELENA facility, the PAX (antiProtonic Atom X-ray spectroscopy) project proposes spectroscopy of Rydberg $\pbar$ atoms using a TES detector to provide tests of bound-state quantum electrodynamics (BSQED) with unprecedented sensitivities~\cite{Paul2021,Baptista2025}.

The target transition of $\pbar$Ca atoms is found  ideal  also in view of TES application. The   $6h\to5g$ transition has an energy around 120~keV with a width of about 35~eV~\cite{Hartmann2002}. This is within the range of a TES with a tin absorber which  is expected to have a resolution around 50 to 70~eV in this region~\cite{Noroozian2013,Saito2025}. Figure~\ref{fig:pbarCa} shows predicted  spectra of $\bar{p}\,^{48}$Ca atoms, highlighting an    improved resolution achievable by a TES detector (50 eV in FWHM)  compared to that of a conventional high-purity germanium (HPGe) detector (700 eV in FWHM). Using an HPGe detector, the PS209 experiment determined the isotope-dependent strong-interaction shifts, with magnitudes ranging from 5 to 33~eV and uncertainties between 10 and 30~eV~\cite{Hartmann2002}. The strong-interaction  width was indirectly estimated from the balance of  intensities within the transition cascade~\cite{Eisenberg1961}. 
A TES detector will allow direct determination of the strong-interaction widths, and provide $\mathcal{O}(1)$~eV precision for the strong-interaction shift measurements.

As seen in Figure~\ref{fig:pbarCa}, the fine structure can be resolved with a TES derector. 
In the analysis by Yoshimura \textit{et al}, the fine-structure separation of 235~eV  between  $6h_{11/2}\to5g_{9/2}$ and $6h_{9/2}\to5g_{7/2}$ was found to be insensitive to the strong interaction~\cite{Yoshimura2024}, therefore this feature can be used to improve  fitting or to check   systematic effects such as the detector calibration.

\section{Conclusion}
In view of recent situations  necessitating  further investigation of antinucleon--nucleus interactions, we discussed the possibility of an improved spectroscopy experiment on $\pbar$ atoms, focusing on the use of calcium isotopes with well-studied  nucleon density distributions. By minimizing uncertainties associated with nucleon density  distributions, this approach aims to provide a more accurate extraction of  $b_0$ and $b_1$,  and refine the $\pbar$--nucleus optical potential model. The proposed  experiment, utilizing a TES detector,  could potentially be conducted as a parasitic experiment of the PAX project in the AD/ELENA facility in the near future.

\section*{Acknowledgments}

We thank M. Doser, D. Jido, N. Kuroda, T. Naito, K.V. Protasov, T. Shima, A. Trzcińska, L. Venturelli, and the PAX collaboration (especially, T. Azuma, T. Hashimoto, B. Ohayon, S. Okada, N. Paul and T.Y. Saito) for  valuable discussions.

This work is supported by  the Institute for Integrated Radiation and Nuclear Science, Kyoto University, the Future Development Funding Program of Kyoto University Research Coordination Alliance, the JST FOREST Program (Grant No. JPMJFR2237),  and the Itoh Science Foundation.










\bibliographystyle{JHEP}
\bibliography{EXA-LEAP2024_proc}

\end{document}